\documentclass{article}
\usepackage{spconf,amsmath,epsfig}
\usepackage{stfloats}
\usepackage{amsmath,amssymb,amsfonts}
\usepackage{algorithmic}

\graphicspath{{figs/}}
\usepackage{booktabs}
\usepackage{etoolbox}

\usepackage{multirow}
\usepackage{siunitx}
\usepackage{cite}
\usepackage{subfig}
\let\OLDthebibliography\thebibliography
\renewcommand\thebibliography[1]{
  \OLDthebibliography{#1}
  \setlength{\parskip}{0pt}
  \setlength{\itemsep}{0pt plus 0.3ex}
}

\begin{document}
	
\title{Color-NeuraCrypt: Privacy-Preserving Color-Image Classification Using Extended Random Neural Networks}
	
\name{
		Zheng Qi, AprilPyone MaungMaung and Hitoshi Kiya}
	
\address{
\begin{tabular}{c}
		    Tokyo Metropolitan University\\
			6-6, Asahigaoka, Hino-shi, Tokyo, 191–0065, Japan\\
			Phone/FAX:+81-042-585-8454\\
			E-mail: \{qi-zheng@ed., apmaung@, kiya@\}tmu.ac.jp
\end{tabular}
}

	
	
	
\maketitle
	
\section*{Abstract}
	
In recent years, with the development of cloud computing platforms, privacy-preserving methods for deep learning have become an urgent problem. NeuraCrypt is a private random neural network for privacy-preserving that allows data owners to encrypt the medical data before the data uploading, and data owners can train and then test their models  in a cloud server with the encrypted data directly. However, we point out that the performance of NeuraCrypt is heavily degraded when using color images. In this paper, we propose a Color-NeuraCrypt to solve this problem. Experiment results show that our proposed Color-NeuraCrypt can achieve a better classification accuracy than the original one and other privacy-preserving methods.
	
\section{Introduction}
In recent years, the spread of deep neural networks (DNNs)~\cite{dnn} has greatly contributed to solving complex tasks for many applications, and it has been very popular for data owners to train DNNs on large amounts of data in cloud servers.
However, data privacy such as personal medical records, may be compromised in that process, because a third party can access the uploaded data illegally, so it is necessary to protect data privacy in cloud environments, and privacy-preserving methods for deep learning have become an urgent challange~\cite{overview}.
One of the most efficient solutions is to encrypt data before the data uploading, so that data owners can train and then test their DNNs in a cloud server with the encrypted data directly~\cite{enc1, enc2,enc3, enc4}.
	
NeuraCrypt~\cite{neura} is a private random neural network that allows us to encrypt data before uploading. Vision Transformation (ViT)~\cite{vit} models have been demonstrated to maintain a high classification performance for medical images (with one channel) under the use of NeuraCrypt, but we point out that the performance of NeuraCrypt is heavily degraded when using color images. In this paper, we extend NeuraCrypt from one channel to three channels, called Color-NeuraCrypt, to avoid performance degradation. Experiment results show that our proposed Color-NeuraCrypt achieved a better classification accuracy than the original one and outperformed other privacy-preserving methods on the CIFAR-10 dataset.

\section{Related Work}
\begin{figure}
		\centerline{\includegraphics[width=\linewidth]{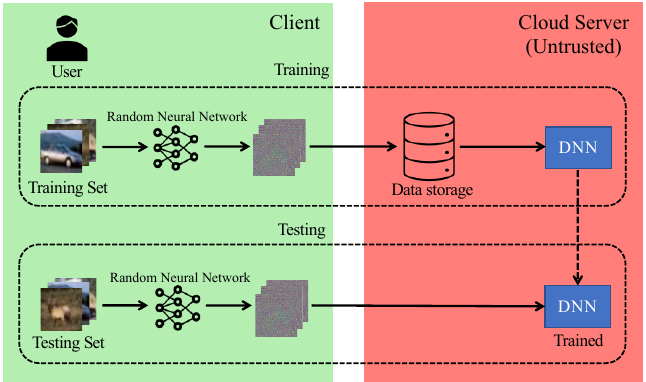}}
		\caption{Framework of proposed method.\label{fig:overview}}
\end{figure}

Lightweight privacy-preserving methods, called learnable encryption, have almost the same usage scenario as the random neural network. Generally, Privacy-preserving image classification methods have to satisfy two requirements: high classification accuracy and strong robustness against various attacks. Tanaka first introduced a block-wise learnable image encryption (LE) method with an adaptation layer~\cite{le}, which is used prior to a classifier to reduce the influence of image encryption. Another encryption method is a pixel-wise encryption (PE) method in which negative-positive transformation and color component shuffling are applied without using any adaptation layer~\cite{pe}. However, both encryption methods are not robust enough against ciphertext-only attacks as in~\cite{itn}. To enhance the security of encryption, LE was extended by adding a block scrambling step and a pixel encryption operation with multiple keys (hereinafter denoted as ELE)~\cite{ele}. However, ELE still has a lower accuracy than that of using plain images. Recently, block-wise learnable encryption methods with an isotropic network have been proposed to reduce the influence of image encryption~\cite{qi,maung}.
Meanwhile, NeuraCrypt was proposed with ViT and achieved a good performance on grayscale medical images, but its performance degraded heavily for color images. In addition, it cannot be directly applied to a standard pre-trained ViT. Accordingly, we propose a novel random neural network called Color-NeuraCypt to improve these issues that the conventional methods have.

\section{Proposed Method}
	\subsection{Overview}
	
	\begin{figure}[t]
		\centering
		\subfloat[]{\includegraphics[width=\linewidth]{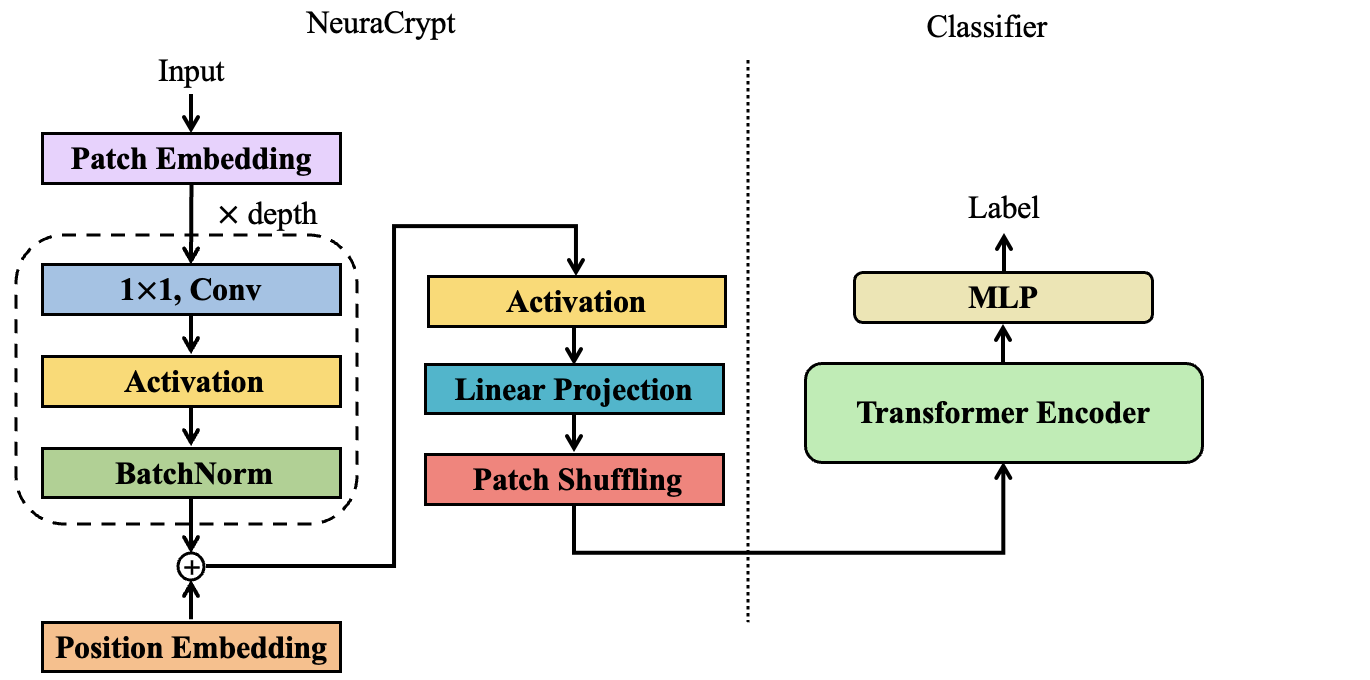}%
			\label{fig:neura}}
		\hfil
		\subfloat[]{\includegraphics[width=\linewidth]{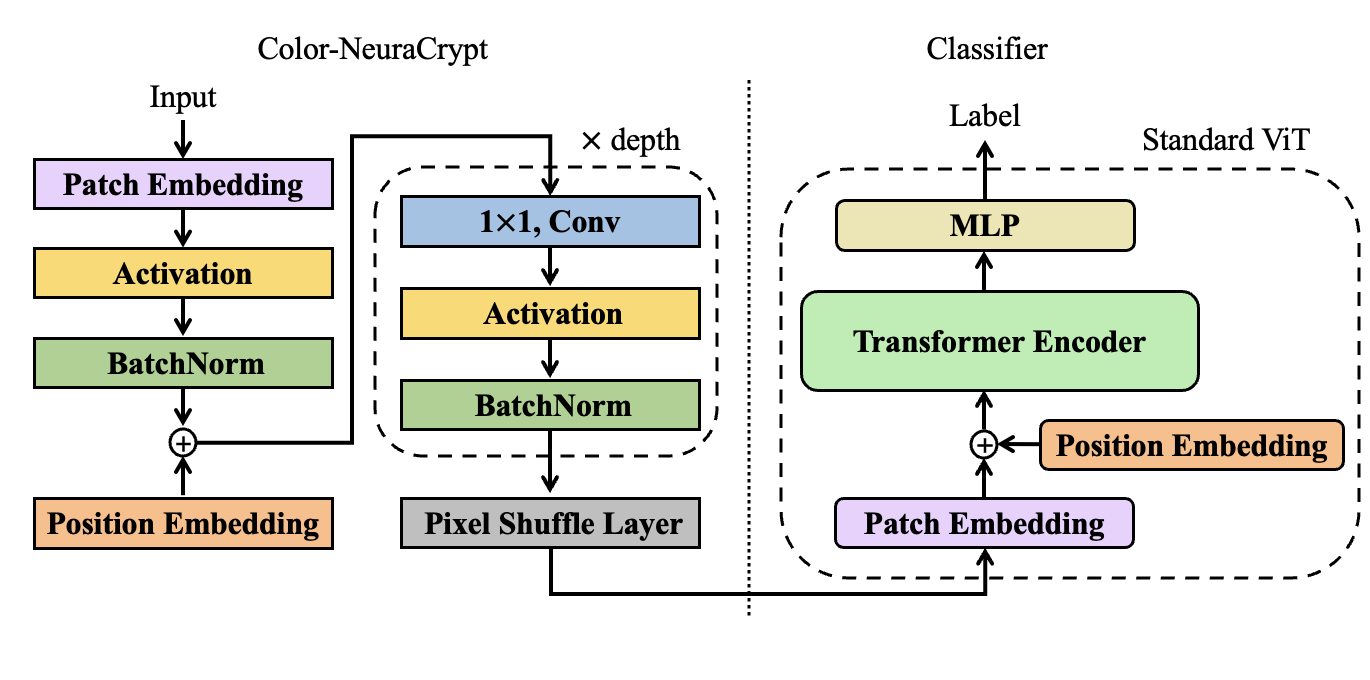}%
			\label{fig:colorneura}}
		\caption{Architecture of two random neural networks (a) NeuraCrypt  (b)  Color-NeuraCrypt (proposed) \label{fig:arc}}
	\end{figure}
	
Figure~\ref{fig:overview} depicts the framework of the proposed scheme. A user encrypts training images by using a random neural network and sending the encrypted images to a cloud provider. Next, the cloud provider trains a ViT model with the uploaded encrypted images without perceiving any visual information. After training, the user also encrypts the testing images using the same random neural network as in training, and sends that to the cloud server. Data privacy can be protected in both training and testing processes in this framework.
	
\subsection{Color-NeuraCrypt}
	
NeuraCrypt is a randomly constructed neural network to encode input data~\cite{neura} as shown in Fig.~\ref{fig:neura}. It consists of patch embedding, several blocks of a $1 \times 1$ convolutional layer, position embedding, and linear projection. It can achieve a high classification accuracy for grayscale medical images, but its performance significantly drops for color images (see Section~\ref{exp}). 
	
\begin{figure}[htbp]
			\begin{minipage}[b]{.48\linewidth}
				\centering
				\centerline{\includegraphics[width=3.5cm]{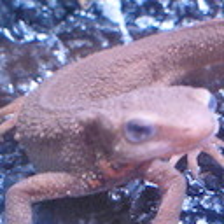}}
				\centerline{(a) Plain}\medskip
			\end{minipage}
			\hfill
			\begin{minipage}[b]{0.48\linewidth}
				\centering
				\centerline{\includegraphics[width=3.5cm]{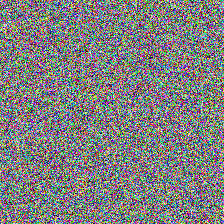}}
				\centerline{(b) Encrypted}\medskip
			\end{minipage}
			\caption{Example of plain and encrypted images.\label{fig:exp}}
		\end{figure}

To avoid performance degradation for color images, we propose a novel random neural network called Color-NeuraCrypt. Figure~\ref{fig:colorneura} shows the architecture of Color-NeuraCrypt. There are two major differences between the two random neural networks:
	
\begin{itemize}
		\item The output of NeuraCrypt is a patch representation. In contrast, the output of Color-NeuraCrypt is an image because we add a pixel shuffling layer at the end of the Color-NeuraCrypt to reshape a patch representation. Figure~\ref{fig:exp} shows an example of plain and encrypted images.
		\item NeuraCrypt randomly permutes patches at the output independently for each image in patch shuffling. In contrast, to align with a standard ViT, we remove the patch shuffling step but still retain the random position embedding to hide the spatial information of plain images.
\end{itemize}
	
Furthermore, we utilize a standard pre-trained ViT, which has trainable patch embedding and position embedding. We fine-tune ViT with encrypted images for training and testing.

		\section{Experiments}
		\label{exp}
		
		\robustify\bfseries
		\sisetup{table-parse-only,detect-weight=true,detect-inline-weight=text,round-mode=places,round-precision=2}
		\begin{table*}[htbp]
			\caption{Comparison of accuracy (\SI{}{\percent}) of image classification. EtC and ELE are block-wise learnable image encryption methods that use a Shakedrop network as a classifier~\cite{shake,ele}, and an adaptation network (denoted as AdaptNet) is applied to reduce the influence of image encryption.\label{tab:results}}
			\centering
			\begin{tabular}{lcccc}
				\toprule
				\multirow{2}{*}{\bfseries Encryption} & \multirow{2}{*}{\bfseries Classifier} &\multirow{2}{*}{\bfseries Image size}& \multicolumn{2}{c}{\bfseries Accuracy} \\
				& && {\bfseries MINST} & {\bfseries CIFAR-10}  \\
				\midrule
				{EtC~\cite{etc,ele}} & AdaptNet+ShakeDrop &$32\times 32$&{--}& 89.09\\
				{ELE~\cite{ele}} &AdaptNet+ShakeDrop &$32\times 32$& {--}& 83.06\\
				\midrule
				{NeuraCrypt~\cite{neura}} & ViT-B\_16 &$224\times 224$ & 97.93 & 70.60 \\
				{Color-NeuraCrypt(proposed)} & ViT-B\_16&$224\times 224$  & {--} & \bfseries \num{96.20}\\
				\midrule
				
				Plain & ShakeDrop &$32\times 32$&{--}  & 96.70 \\
				Plain & ViT-B\_16 & $224\times 224$  & 99.71 &{99.10}\\
				\bottomrule

			\end{tabular}
		\end{table*}
		
We conducted image classification experiments on the MNIST~\cite{minst} and CIFAR-10~\cite{cifar10} datasets. The MNIST dataset consists of 70,000 grayscale images  (dimension of $1 \times 28 \times 28$) of handwritten digits with ten classes, where 60,000 images are for training and 10,000 for testing. The CIFAR-10 dataset consists of 60,000 color images (dimension of $1 \times 28 \times 28$), where 50,000 images are for training and 10,000 for testing.
		
We used a PyTorch implementation of ViT\footnote{https://github.com/jeonsworld/ViT-pytorch} and fine-tuned the ViT-B\_16 model which was pre-trained with the ImageNet21k dataset. To maximize the classification performance, we followed the training settings from~\cite{vit} except for the learning rate. The parameters of the stochastic gradient descent (SGD) optimizer for encrypted images that we used were: a momentum of 0.9, a weight decay of 0.0005, and a learning rate value of 0.03-0.1. In addition, the depth of NeuraCrypt and Color-NeuraCrypt was set to 4.
		
As shown in Table~\ref{tab:results}, ViT models with NeuraCrypt performed with 97.93\% accuracy on the MNIST dataset, which was similar to medical images. However, it achieved unsatisfactory accuracy on the CIFAR-10 dataset. These results confirmed that NeuraCrypt is very effective on grayscale images but difficult to be applied to color images.
Table~\ref{tab:results} also shows the classification performance of other privacy-preserving methods. Our Color-NeuraCrypt outperformed not only NeuraCrypt but also the two block-wise encryption methods (ELE and EtC~\cite{etc}) on the CIFAR-10 dataset, so the proposed method was confirmed to be more suitable for color images.
		
\section{Conclusion and Future Work}
In this research, we proposed a random neural network, called Color-NeuraCrypt for privacy-preserving. Color images encrypted by Color-NeuraCrypt can be applied to ViT models for both training and testing directly. Experiment results showed that our Color-NeuraCrypt achieved a better accuracy than NeuraCrypt and other privacy-preserving methods on color images.
		
As a random neural network is considered as an encryption method for privacy-preserving, its security needs to be evaluated. For example, we can correctly match a plain and encrypted sample using the algorithm in~\cite{nopri}. Furthermore, a random neural network can hide the visual information of plain images, but it is hard to secrete some transparent information, such as the distribution of the dataset and the encryption scheme. An attacker may perform a ciphertext-only attack via that information to reconstruct visual information from encrypted images.
		
\section*{{\hfil Acknowledgment}}
		This study was partially supported by JSPS KAKENHI (Grant Number JP21H01327).

\bibliographystyle{IEEEtran}
\bibliography{refs}

\end{document}